\documentclass[aps,prx, superscriptaddress, showpacs,twocolumn,nofootinbib,longbibliography,nobalancelastpage]{revtex4-1}

\usepackage{graphicx, color}
\usepackage{dcolumn}
\usepackage{bm}
\usepackage{amsmath}
\usepackage{amssymb}
\usepackage{verbatim}
\usepackage{appendix}
\usepackage{float}
\usepackage{color, soul}
\usepackage{gensymb}

\newcommand{\beginsupplement}{%
        \setcounter{table}{0}
        \renewcommand{\thetable}{S\arabic{table}}%
        \setcounter{figure}{0}
        \renewcommand{\thefigure}{S\arabic{figure}}%
             \setcounter{equation}{0}
        \renewcommand{\theequation}{S\arabic{equation}}%
     }

\begin{document}

\title{Molecular Parity Nonconservation in Nuclear Spin Couplings}

\author{John W. Blanchard}
\email{blanchard@uni-mainz.de}
\affiliation{Helmholtz-Institut Mainz, GSI Helmholtzzentrum f{\"u}r Schwerionenforschung GmbH, 55128 Mainz, Germany}

\author{Jonathan P. King}
\affiliation{Department of Chemistry, University of California at Berkeley, CA, 94720 USA}
\affiliation{Materials Science Division, Lawrence Berkeley National Laboratory, Berkeley, CA, 94720 USA}

\author{Tobias F. Sjolander}
\affiliation{Department of Chemistry, University of California at Berkeley, CA, 94720 USA}
\affiliation{Materials Science Division, Lawrence Berkeley National Laboratory, Berkeley, CA, 94720 USA}

\author{Mikhail G. Kozlov}
\affiliation{Petersburg Nuclear Physics Institute, Gatchina 188300, Russia}
\affiliation{St. Petersburg Electrotechnical University ``LETI'', St. Petersburg 197376 Russia}

\author{Dmitry Budker}
\affiliation{Helmholtz-Institut Mainz, GSI Helmholtzzentrum f{\"u}r Schwerionenforschung GmbH, 55128 Mainz, Germany}
\affiliation{Johannes Gutenberg-Universit{\"a}t Mainz, 55099 Mainz, Germany}
\affiliation{Department of Physics, University of California, Berkeley, CA 94720-7300 USA}


\date{\today}

\begin{abstract}
The weak interaction does not conserve parity, which is apparent in many nuclear and atomic phenomena. However, thus far, parity nonconservation has not been observed in molecules. Here we consider nuclear-spin-dependent parity nonconserving contributions to the molecular Hamiltonian. These contributions give rise to a parity nonconserving indirect nuclear spin-spin coupling which can be distinguished from parity conserving interactions in molecules of appropriate symmetry, including diatomic molecules. We estimate the magnitude of the coupling, taking into account relativistic corrections. Finally, we propose and simulate an experiment to detect the parity nonconserving coupling using liquid- or gas-state zero-field nuclear magnetic resonance of electrically oriented molecules and show that $^{1}$H$^{19}$F should give signals within the detection limits of current atomic vapor-cell magnetometers.
\end{abstract}


\maketitle

\section{Introduction}

Parity nonconservation (PNC) in the weak interaction was first theorized \cite{Lee1956} in 1956, followed by the first experimental observation in $\mathrm{\beta}$-decay of $^{60}$Co nuclei in 1957 \cite{Wu1957}. 
In the decades since, a variety of parity nonconserving effects have been observed in atoms \cite{Bouchiat2012,SBDJ18}. 
PNC should also be present in the molecular Hamiltonian \cite{Ginges2004,Berger2019}, although its effects have not yet been observed. 
Molecules afford some interesting possibilities to observe PNC effects including the proposed detection of energy shifts between enantiomers of chiral molecules \cite{GaMa74a,GaMa74,Letokhov75,Kompanet1976,HarSto80,GKL82e,Bauder1997,Lahamer2000,Darquie2010,Crassous2003,Cournol2019}, time-dependent optical activity in chiral molecules \cite{HarSto78,Nagy1999,MacDermott2004}, and Stark interference in linear di- or polyatomic molecules \cite{DeMille2008,Cahn2014,AACD18,Cournol2019}. 
Molecules may also provide opportunities to study finer aspects of PNC. 
In particular, molecules are of interest because of the presence of closely lying levels of opposite parity which are not a general feature of atoms other than hydrogen \cite{Flambaum1985,Kozlov1995,Khr91}. 



Here we propose to observe nuclear-spin dependent PNC via the indirect nuclear spin-spin coupling in molecules. 
It has been known for some time that PNC effects should cause frequency shifts between the nuclear magnetic resonance (NMR) spectra of enantiomers of chiral molecules \cite{GKL82e,BaRo96,RobBar01,LB03,WMV05,LB06,Bast2006,WBMSV07,NaBe09,Nahrwold2014,Eills2017}. 
Here we consider a different effect which can be observed in non-chiral molecules, including diatomic molecules. 
The indirect nuclear spin-spin coupling (\textit{J}-coupling) is a commonly measured property in nuclear magnetic resonance spectroscopy \cite{Gutowsky1951,Hahn1952,Ramsey1953}. 
The Hamiltonian for two coupled nuclei is bilinear with respect to their spin vector operators ${\bm{I}}$ and ${\bm{S}}$. 
The isotropic or scalar \textit{J}-coupling, parametrized by a tensor transforming under rotations as a rank-0 spherical tensor, is most commonly observed in liquid-state NMR spectroscopy. 
The corresponding Hamiltonian is:
\begin{equation}
{H_0}=2\pi J{\bm{I}}\cdot {\bm{S}}.
\label{scalar}
\end{equation}
In general, the \textit{J}-coupling interaction is characterized by a reducible rank-two tensor $J_{ik}$:
\begin{equation}
{H_{J}}=2\pi {I_i J_{ik} S_k}.
\label{tensor}
\end{equation}
Nonzero-rank contributions to the \textit{J}-coupling can be observed in oriented molecules, for example a rank-one coupling can be written as
\begin{equation}
{H_1}=2\pi\bm{J}^{(1)}\cdot {\bm{I}}\times {\bm{S}},
\label{rank1}
\end{equation}
where $J^{(1)}_i=\epsilon_{ijk}J_{jk}$ is a (pseudo)vector when the interaction is P-odd(even)  
and $\epsilon_{ijk}$ is the Levi-Civita tensor. 
The effects of ${H_1}$ have not yet been observed since it averages to zero in isotropically rotating molecules and is suppressed in a magnetic field if spins $I$ and $S$ have different Larmor frequencies. 
It could, in principle, be observed in solids, but solid-state NMR typically suffers from low resolution. 
Zero-field nuclear magnetic resonance of electrically oriented molecules can reveal signals from ${H_1}$, as was proposed for absolute determination of molecular chirality \cite{GaBu16,King2017}. 
Here, we propose to apply this technique to observe the interaction ${H_1}$ in a non-chiral molecule, where it can appear due to the contribution of the nuclear-spin-dependent parity nonconserving weak interaction \cite{Barra1996}.  

For oriented molecules, the residual component of $\bm{J}^{(1)}$ must be parallel to the orientation axis. 
A parity nonconserving $\bm{J}^{(1)}$ with nonzero projection along the molecular electric dipole can be found in molecules belonging to several symmetry point groups (see Supplemental Material) including diatomics ($C_{\infty v}$). 
Of course, diatomics are non-chiral and, therefore, there can be no observable parity conserving contribution to $\bm{J}^{(1)}$.

There is a similarity between the effect considered here and the hyperfine correction to the nuclear-spin-independent PNC interaction, which mimics the nuclear-spin-dependent PNC interaction \cite{FK85a,BP91b}. 
There both interactions are centered on the same nucleus. 
Since these interactions diverge at the origin, one ends up with a highly singular effective operator, which is difficult to calculate accurately. 
Here we take the hyperfine interaction and nuclear-spin-dependent PNC interaction on different nuclei. 
This gives us the PNC spin-spin coupling, which is described by a less singular effective operator that may be easier to calculate (see below).

\section{Theory}

We now estimate the magnitude of $\bm{J}^{(1)}_\mathrm{PNC}$ in the diatomics $^{205}$Tl$^{19}$F and $^{1}$H$^{19}$F.  
$\bm{J}^{(1)}_\mathrm{PNC}$ arises from the nuclear-spin-dependent parity-odd weak interaction. 
For a first-order approximation, we only include terms that are linear in the nuclear spin operators $\bm{{I}}_K$ (we use the notation $\bm{{I}}_K$ for a generic nuclear spin operator, but write $\bm{{I}}$ and $\bm{{S}}$ when explicitly considering a two-spin Hamiltonian for calculation of observables in an NMR experiment). We also neglect terms that contain electron spin operators since, to a first-order approximation, they do not contribute to PNC in the diamagnetic molecules under consideration \cite{Khr91}. 

In the non-relativistic approximation the Hamiltonian for the $\bm{J}^{(1)}$ parity nonconserving interaction has the form (using atomic units $\hbar=m_e=|e|=1$):
 \begin{align}\label{NSD-PNC}
 {H}_\mathrm{PNC}
 &=\frac{G\alpha}{2\sqrt2}
 \sum_{i,K} g_K \bm {\bm{I}}_K
 \left[\bm p_i,\delta(\bm r_i-\bm R_K)\right]_+\,,
 \end{align}
where $G\approx2.22\times 10^{-14}$ is the Fermi constant, $\alpha\approx 1/137$ is the fine-structure constant, $\bm p_i=-i\bm{\nabla}_i$ and $\bm r_i$ are the momentum and coordinate of the $i$-th electron, and $\bm R_K$ is the coordinate of the nucleus $K$. The dimensionless coupling constant $g_K$ is of order unity. Two main contributions to this coupling come from the nuclear anapole moment \cite{FK80} and the electroweak electron vector- and nucleon axial-vector interaction \cite{Khr91}. For the heavy nuclei the anapole contribution usually dominates \cite{FKS84a,FM97,DeMille2008}.

To account for a magnetic field, we substitute the canonical momentum \cite{ASVJ99}  
\begin{equation}
\bm p_i \rightarrow \bm \pi_i = \bm p_i +\alpha \bm A, \label{Eq:canonicalp}
\end{equation}
where $\bm A$ is vector potential. 
In the case of spin-spin coupling, the
magnetic field is produced by the magnetic moment of nucleus $L$: $\bm
\mu_L=\gamma_L\bm I_L$, where $\gamma_L$ is the gyromagnetic ratio.
We can then take:
 \begin{align}\label{vec_pot}
 {\bm A}_L
 &=\gamma_L\frac{{\bm I}_L\times (\bm r -\bm R_L)}
 {|\bm r -\bm R_L|^3}\,.
 \end{align}
Substituting Eqs.\,\eqref{vec_pot} and \eqref{Eq:canonicalp} in Eq.\,\eqref{NSD-PNC} we obtain:
\begin{widetext}
 \begin{align}\label{NSD-PNC-A}
 {H}_\mathrm{PNC}
 &=\frac{G\alpha}{2\sqrt2}
 \sum_{i,K} g_K {\bm I}_K
 \left[\bm p_i,\delta(\bm r_i-\bm R_K)\right]_+
 +\frac{G\alpha^2}{\sqrt2}
 \sum_{i,K,L} g_K \gamma_L
 \frac{{\bm I}_K \cdot ({\bm I}_L\times (\bm R_K-\bm R_L))}
 {|\bm R_K-\bm R_L|^3}
 \delta(\bm r_i-\bm R_K)
 \,.
 \end{align}
 The second term here is bilinear in nuclear spin operators and contributes to the spin-spin coupling \cite{Barra1996}. 
 By comparison to Eq.\,\eqref{rank1} we obtain $ \bm{J}^{(1)}_\mathrm{PNC}$ in vector form:
 \begin{align}\label{mixed-PNC}
 \bm{J}^{(1)}_\mathrm{PNC}
 &=\frac{G\alpha^2}{2\pi\sqrt2}
 \left(\gamma_S g_I
 \langle \Psi_e|\sum_{i}\delta(\bm r_i-\bm R_I)|\Psi_e\rangle
 - g_S \gamma_I
 \langle \Psi_e|\sum_{i}\delta(\bm r_i-\bm R_S)|\Psi_e\rangle
 \right)
 \frac{\bm R_I-\bm R_S}
 {|\bm R_I-\bm R_S|^3}
 \,.
 \end{align}
\end{widetext}

The electronic matrix elements in \eqref{mixed-PNC}
correspond to the total electronic densities on the respective nuclei:
 \begin{align}\label{ME_delta}
 \langle \Psi_e|\sum_{i}\delta(\bm r_i-\bm R_K)|\Psi_e\rangle
 &= \rho_e(\bm R_K)\,.
 \end{align}
Substituting into (\ref{mixed-PNC}) we obtain:
 \begin{align}\label{mixed-PNCa}
 \bm{J}^{(1)}_\mathrm{PNC}
 =\frac{G\alpha^2}{2\pi\sqrt2}
 \bigg(\gamma_S g_I
 \rho_e(\bm R_I)
 - g_S \gamma_I
 \rho_e(\bm R_S)
 \bigg)\frac{\bm R_{I,S}}
 {|\bm R_{I,S}|^3}
 \,,
 \end{align}
where $\bm R_{I,S}= \bm R_I -\bm R_S$.
The dominant contribution to the density \eqref{ME_delta} comes from
the two K-shell electrons, whose wavefunctions are
hydrogenic \cite{Sob79}, $\psi_{1s}\approx 2\sqrt{Z^{3}/4\pi}\,\mathrm{e}^{-Zr}$,
therefore
 \begin{align}\label{e-density}
 \rho_e(\bm R_K)\approx \frac{2}{\pi}Z^3_K\,.
 \end{align}
The contribution of the $2s$ shell is approximately eight times smaller and can be neglected in the estimates (for hydrogen, the density at the origin scales as $1/n^3$). Putting \eqref{e-density} in \eqref{mixed-PNCa} we arrive at
 \begin{align}\label{mixed-PNCb}
 \bm{J}^{(1)}_\mathrm{PNC}
 =\frac{G\alpha^2}{\pi^2\sqrt2}
 \left(\gamma_S g_I Z_I^3
 - g_S \gamma_I Z_S^3
 \right)\frac{\bm R_{I,S}}
 {|\bm R_{I,S}|^3}
 \,.
 \end{align}

Typical internuclear distances $R_{I,S}$ are comparable to the bond length and are about 3 --- 4 Bohr radii. 
Given the $Z^3$ scaling, the term in parentheses in Eq.\,\eqref{mixed-PNCa} including $Z$ from the heaviest atom will dominate. 
Assuming $R_{I,S}\approx 4$, the magnitude of $\bm{J}^{(1)}_\mathrm{PNC}$ is estimated:
 \begin{equation}\label{J_estimate}
 J^{(1)}_\mathrm{PNC}\sim \frac{G\alpha^2}{16\pi^2\sqrt2}\gamma_S g_I Z_I^3,
 \end{equation}
where $Z_I$ is the charge of the heaviest nucleus in the molecule.

For heavier atoms, relativistic effects become important. The relativistic enhancement factor is (Supplemental Material) \cite{Bouchiat1974,Khr91}:
 \begin{align}\label{RelEnhancement}
 &F_\mathrm{rel}
 = \frac{2(1+\gamma)(2ZR_\mathrm{nuc})^{2\gamma-2}}
 {\Gamma^2(2\gamma+1)}\,,
 \,\,
 \gamma=\sqrt{1-(\alpha Z)^2},
 \end{align}
where $R_\mathrm{nuc}$ is the nuclear radius in atomic units and $\Gamma$ refers to a gamma function. Equation \eqref{mixed-PNCb} becomes:
 \begin{align}\label{Eq:final_PNC}
 \bm{J}^{(1)}_\mathrm{PNC}
 =\frac{G\alpha^2}{\pi^2\sqrt2}
 \left(\gamma_S g_I Z_I^3 F_{\mathrm{rel},I}
 - g_S \gamma_I Z_S^3 F_{\mathrm{rel},S}
 \right)\frac{\bm R_{I,S}}
 {|\bm R_{I,S}|^3}
 \,.
 \end{align}
 

 \subsection{Estimates}

 Let us estimate the magnitude $J^{(1)}_\mathrm{PNC}$ for the molecules $^{205}$Tl$^{19}$F and $^{1}$H$^{19}$F. 
 TlF is a popular candidate for molecular tests of fundamental symmetries because the heavy $^{205}$Tl atom (Z=81) is expected to give rise to strong P- and T-violating effects \cite{HS80,CSH91,VGE82e,NEMS17,SBDJ18}. 
 From Eq.\ \eqref{Eq:final_PNC} neglecting contribution of the fluorine and using internuclear distance $R=3.93$ a.u.\ and relativistic factor $F_\mathrm{rel,Tl}=7.6$ we obtain:
 \begin{equation}\label{estimate_TlF}
 J^{(1)}_{\mathrm{PNC,TlF}}\approx 9\times 10^{-3} g_\mathrm{Tl}\,\mathrm{Hz}.
 \end{equation}
 According to \citet{FKS84a} the coupling constant for Tl is $g_\mathrm{Tl}\approx 0.5$.
 While this value is promising given the resolution of zero-field NMR, diatomic TlF does not exist in the liquid phase and zero-field NMR would be a significant experimental challenge. 
 A molecular-beam experiment featuring electric-field orientation could be a good option for TlF \cite{Gra2019TlF}.
 Indeed, hyperfine measurements with molecular beams have been used to determine the scalar and symmetric rank-2 components of the TlF \textit{J}-coupling \cite{Boeckh1964,Bryce2000}.
 
 
 \begin{figure}
 \includegraphics[width=\columnwidth]{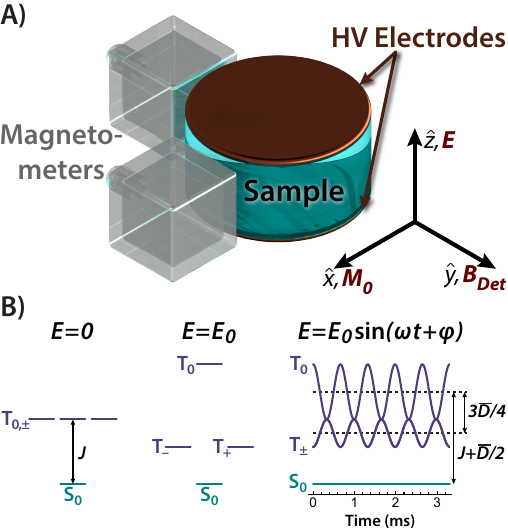}%
 \caption{\textbf{Experimental arrangement:}
 A) The apparatus includes a sample made up of polar achiral molecules, each of which is a two-spin heteronuclear spin system, oriented by an electric field $\bm{E}$ produced by high-voltage (HV) electrodes.
 The nuclear spin state is prepared such that the initial magnetization is oriented along $\bm{M_0}$, and magnetometers (depicted here as glass vapor cells) are used to measure the nuclear spin magnetization projection along $\bm{B_{Det}}$. 
 B) Nuclear spin eigenstates for various electric-field conditions. In the absence of any applied fields are three degenerate triplet states and a singlet state, separated by the $J-$coupling.
 Application of a dc electric field introduces a residual dipole-dipole coupling of strength $\overline{D}$ that shifts the $\left|T_0 \right\rangle$ state by $\overline{D}$ and the $\left|T_\pm \right\rangle$ states by $-\overline{D}/2$.
 An ac electric field modulates the triplet states such that the average $\left|S_0 \right\rangle \leftrightarrow \left|T_0 \right\rangle$ separation is $J+\overline{D}/2$, and the average $\left|T_0 \right\rangle \leftrightarrow \left|T_\pm \right\rangle$ separation is $3\overline{D}/4$.
 }
 \label{Fig:Experimental}
 \end{figure}
 
 Let us estimate the PNC coupling for $^{1}$H$^{19}$F. 
 The relativistic factor \eqref{RelEnhancement} for fluorine is very close to unity, so we can use the non-relativistic expression \eqref{mixed-PNCb}, which gives:
 \begin{equation}\label{estimate_HF}
 J^{(1)}_{\mathrm{PNC,HF}}\approx 9\times 10^{-6} g_\mathrm{F}\, \mathrm{Hz}.
 \end{equation}
 Liquid HF would be an option, as the neat liquid has a concentration of $\sim$50~M at its boiling point (19.5\degree C), both ${}^1$H and $^{19}$F are naturally abundant spin-1/2 isotopes, and the molecular dipole moment of 1.86~D is reasonably large (a large dipole moment is needed to achieve a high degree of molecular orientation).
 Unfortunately, due to hydrogen-bonding and autoionization effects, $J_{\mathrm{HF}}$ can only be observed when HF is diluted in aprotic solvents or when the neat liquid is cooled to very low temperatures \cite{Shamir1973,Martin1974,Mackor1968}.
 
 %
 %
 %
A possible alternative could be the endofullerene complex HF@C$_{60}$ \cite{krachmalnicoff2016}, though the molecular dipole moment is decreased to 0.45~D, and in order to avoid broadening due to intermolecular dipole-dipole couplings, it would likely be necessary to dissolve the HF@C$_{60}$ in a solvent such as 1-chloronaphthalene -- this would result in an overall decrease in concentration by some three orders of magnitude.
More suitable from the experimental standpoint may be liquid or gaseous fluoromethane (CH$_3$F), which has a molecular dipole moment of 1.81~D and a concentration of $\sim$16~M at its saturation pressure of 33~bar at $25^\circ$C. 
The magnitude of the PNC $J$-couplings for this molecule is estimated in the microhertz range\footnote{
The CH$_3$F molecule can have three different couplings, HC, HF, and CF. For the HF coupling we can use the same estimate as for the HF molecule and scale it by the square of the distance $R_{HF}$, which is here 3.8\,au instead of 1.7\,au. This gives us 9\,$\mu$Hz$\cdot(1.7/3.8)^2\approx 2\,\mu$Hz. For the HC coupling in addition to the distance (2.05) we need to change the charge, so we get: $9\,\mu$Hz$\cdot(1.7/2.1)^2\cdot(6/9)^3 \approx 2\,\mu$Hz. Thus, both couplings are of the same size and about five times smaller than in HF. The coupling of C and F is different. Here we have both terms in Eq.\,\eqref{Eq:final_PNC} and they almost cancel each other, so the result is much smaller and unreliable. 
If we polarize the molecule in an electric field, we need to project the vector $\bm{J}^{(1)}$ on the molecular axis. This will give us additional suppression by $\cos\phi$, where $\phi$ is the angle between molecular axis and $\bm{J}^{(1)}$. For HC $\cos\phi\approx$0.3 and for HF $\cos\,\phi\approx$0.8. All in all, we get close to $1\,\mu$Hz and $2\,\mu$Hz for HC and HF PNC couplings, respectively.
}.

 
 $\bm{J}^{(1)}_\mathrm{PNC}$ is fixed with respect to the molecular orientation, but small molecules in the liquid or gas state undergo fast molecular rotation. In NMR, we measure the averaged $\overline{\bm{J}^{(1)}_\mathrm{PNC}}$ which is nonzero only when the molecules are oriented.
The residual rank-1 coupling for a molecule oriented in an electric field is \cite{Buckingham1963}:
\begin{equation}\label{sup:orient}
\overline{J^{(1)}} = \frac{d E}{3kT}J^{(1)},
\end{equation}
 where $d$ is the electric dipole moment, $E$ is the electric field experienced by the sample, $k$ is the Boltzmann constant, and $T$ the temperature. The scaling factor is thus given by $d E/3kT$. In polar liquids $E$ is related to the macroscopic electric field as $E \approx [(\epsilon_r + 2)/3]E_{\mathrm{mac}}$, where $\epsilon_r$ is the static relative permittivity of the medium. For HF, $d$ = 1.86 D, and $\epsilon_r$ = 84. Assuming a temperature of 300 K and $E_{\mathrm{mac}}\approx70\,$kV/cm \cite{Riley2000}, the orientation scaling factor is 0.1.
This gives $\overline{J^{(1)}_\mathrm{PNC,HF}} \approx 1\times10^{-6}$ Hz for the estimated projection of $\overline{\bm{J}^{(1)}_\mathrm{PNC,HF}}$ along the z-axis. 

\subsection{Spin Dynamics}
Zero-field NMR involves measuring the evolution of nuclear magnetism of coupled spins in the absence of external magnetic fields \cite{Blanchard2017} (Fig.\ \ref{Fig:Experimental}a). Signals are typically detected with an atomic vapor-cell magnetometer.  When a diatomic molecule is oriented along the z-axis, the nuclear spin Hamiltonian is:
\begin{multline}
\frac{H}{2\pi}=J{\bm{I}}\cdot {\bm{S}}+\frac{i\overline{J^{(1)}_z}}{2}({{I}^+}{{S}^-}-{{I}^-}{{S}^+})\\
-\overline{D}\left(3I_zS_z-\bm{I}\cdot \bm{S}\right),
\end{multline}
where the residual dipolar coupling \cite{Buckingham1963} is
\begin{equation}
\label{Eq:AverageD}
    \overline{D}=
    \frac{\gamma_I\gamma_S}{r^3}\frac{1}{30}\left( \frac{d E}{k T} \right)^2\,.
\end{equation}
Rapid motion around the z-axis averages components of $\overline{\bm{J}^{(1)}}$ and $\overline{D}$ orthogonal to the z-axis to zero. 
$J$ is the scalar component of the \textit{J}-coupling which is not affected by molecular rotation. 
Even though $\overline{D}$ scales quadratically with the electric field, it cannot be neglected, as the dipolar coupling for an HF molecule fully aligned along $z$ is approximately 150~kHz.
For the conditions described above, the residual coupling $\overline{D}=452~\rm{Hz}$, which is comparable to the isotropic $J$ coupling, $\sim$530~Hz \cite{Muenter1970}.
The $z$-component of the symmetric anisotropic $J$ coupling has a form identical to that of the dipolar coupling and effectively adds to $\overline{D}$, but it is a small effect that can be neglected for the present analysis.


In the absence of electric fields, the nuclear spin eigenstates are the singlet ($|S\rangle$) and three degenerate triplet states ($|T_{0,\pm1}\rangle$), with an energy separation equal to $J$.
The dipolar coupling shifts the $|T_{0}\rangle$ state by $\overline{D}$ and the $|T_{\pm1}\rangle$ states by $-\overline{D}/2$ \cite{Blanchard2015}.


If an oscillating electric field at a frequency $\omega$ is applied, the term proportional to $J_\mathrm{PNC}$ is modulated at this frequency, and the term proportional to $\overline{D}$ is modulated at twice this frequency:
\begin{multline}
\frac{H}{2\pi}=J\,{\bm{I}}\cdot {\bm{S}}
+ \frac{i\sin(\omega t+\phi) \overline{J_\mathrm{PNC}}}{2}({{I}^+}{{S}^-}-{{I}^-}{{S}^+})\\
-\frac{1-\cos(2\omega t+2\phi)}{2}\overline{D}\left(3I_zS_z-\bm{I}\cdot \bm{S}\right)
,
\end{multline}
where $\phi$ is the phase of the AC electric field and $\overline{J_\mathrm{PNC}}$ and $\overline{D}$ now refer to the peak values.
Gathering terms based on time dependence, we find
 \begin{multline}
\frac{H}{2\pi}=\left(J+\frac{\overline{D}}{2}\right){\bm{I}}\cdot {\bm{S}} - \frac{3\overline{D}}{2}I_zS_z\\
+ \frac{i\sin(\omega t+\phi) \overline{J_\mathrm{PNC}}}{2}({{I}^+}{{S}^-}-{{I}^-}{{S}^+})\\
-\frac{\cos(2\omega t+2\phi)}{2}\overline{D}\left(3I_zS_z-\bm{I}\cdot \bm{S}\right)\label{Eq:combined Ham}
.
\end{multline}

\begin{figure*}[hbt]
 \includegraphics[width=175mm]{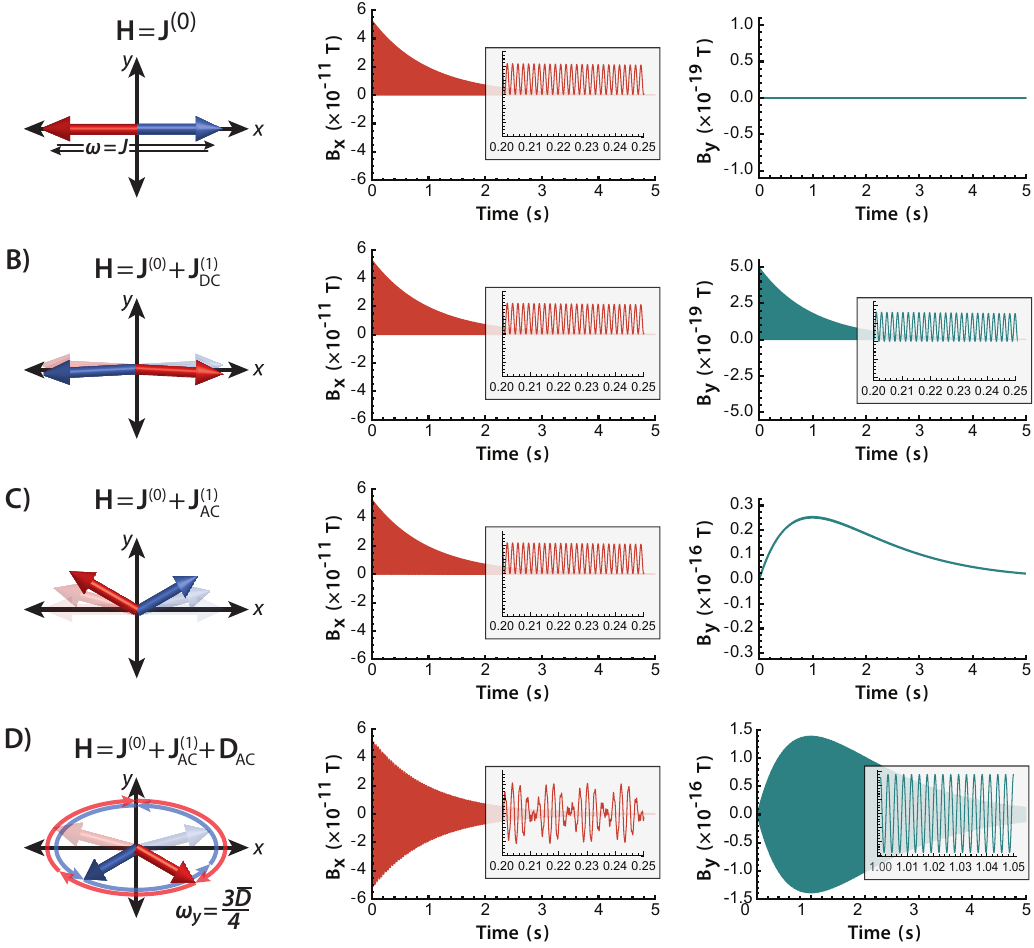}
 \caption{\textbf{Evolution of nuclear magnetization at zero magnetic field:}
 Vector diagrams show the evolution of $\bm{I}$ and $\bm{S}$ magnetization, starting from $I_x-S_x$. Simulated signals along $\hat{x}$ and $\hat{y}$, assuming an initial state proportional to $\gamma_I I_x- \gamma_S S_x$.
 A) Evolution under the isotropic $J$-coupling.
 B) Including a static $\overline{J^{(1)}_z}$.
 C) With a resonantly modulated $\overline{J^{(1)}_z}$.
 D) Including the residual dipole-dipole coupling due to alignment by the electric field. 
 In all cases, the spins evolve only in the $\hat{x}\hat{y}$ plane.
 }
\label{Fig:Evolution}
 \end{figure*}

To gain a physical intuition regarding this system, it is helpful to first consider the case of no interaction other than the scalar $J$-coupling. Suppose we begin by preparing an initial state in which the spins $I$ and $S$ have opposite average projection onto some direction ($\hat{x}$) corresponding to magnetization $\bm{M}_0$ along this direction (the sample is at zero external magnetic field), see Fig.\,\ref{Fig:Evolution}A. 

The effect of the $J$-coupling (i.e., the $\bm{I}\cdot\bm{S}$ interaction) is to produce beats of the spins so that the average values of $\bm{I}$ and $\bm{S}$ oscillate between positive and negative value and the magnetization $\bm{M}$ correspondingly oscillates along $\hat{x}$. At the times when the average values of $\bm{I}$, $\bm{S}$, and $\bm{M}$ go through zero, the polarization of the spin system has higher order and the evolution is akin the process of alignment-to-orientation conversion known in atomic and molecular physics \cite{Auzinsh2010}.
Interpreting the evolution in the product-operator formalism \cite{Keeler2011}, the isotropic $J$-coupling converts $I_x-S_x$ into $I_y S_z - I_z S_y$, then into $S_x-I_x$, etc., as depicted in Fig.\,\ref{Fig:Evolution}A. 

Now let us add an electric field $\vec{E}$ applied along $\hat{z}$ perpendicular to $\bm{M}_0$, which orients the molecules, giving rise to nonzero $\overline{\bm{J}^{(1)}}$ (see Fig.\,1a). 
In the absence of other interactions, the effect of $\overline{\bm{J}^{(1)}}$ would be to convert $I_x-S_x$ into $I_y S_z + I_z S_y$, then into $S_x-I_x$, etc., which would not produce any magnetization effects distinguishable from those of isotropic $J$.

An ``interesting'' effect of $\overline{\bm{J}^{(1)}}$ appears when this coupling is combined with the scalar part to convert $I_y S_z + I_z S_y$ generated by the latter into $I_y+S_y$ associated with a detectable magnetization along $\hat{y}$ (see Fig.\,\ref{Fig:Evolution}B and Appendix for details). 

The essential observable phenomenon revealing the rank-one $J-$coupling in the system is the component of the magnetization perpendicular to both $\overline{\bm{J}^{(1)}}$ and the original direction of the magnetization. 
In fundamental physics experiments, it is customary to consider a rotational invariant, a (pseudo)scalar  built out of experimental parameters that reveals the symmetry of the effect and the behavior of the system under the reversals of the parameters. In this case, we can write such invariant as
\begin{equation}
    \bm{B}_{Det}\cdot \bm{M}_0\times \bm{E},\label{Eq:RotInvariant}
\end{equation}
which is P-odd and T-even (i.e., a pseudoscalar invariant with respect to time reversal), revealing that the relevant signal should reverse sign under the reversals of each of the quantities  $\bm{B}_{Det}$, $\bm{M}_0$, and  $\bm{E}$.

We now discuss the time dependence of the signal that can be evaluated based on Eq.\,\eqref{Eq:combined Ham} and the initial conditions. Let us first consider just the scalar $J-$coupling and neglect the other coupling. As illustrated in Fig.\,\ref{Fig:Evolution}A, there is only magnetization along the initial direction which undergoes oscillations at frequency $J$. Adding the antisymmetric $J-$coupling, we see that there appears a transverse component of magnetization along the detection direction oscillating at the same frequency Fig.\,\ref{Fig:Evolution}B.
The maximum transverse magnetization can be roughly estimated as $M_0\overline{J^{(1)}}/J$.

The effect may be considerably enhanced, by a factor on the order of $JT_2$ if we reverse the direction of the electric field synchronously with the reversals of the main component of the magnetization, Fig.\,\ref{Fig:Evolution}C. 



Finally, we re-introduce the dipole-dipole coupling. The $\bm{I}\cdot\bm{S}$ part of the interaction [see Eq.\,\eqref{Eq:combined Ham}] adds to the scalar $J$-coupling and does not bring any qualitative new features except for changing the frequency of the oscillation of the main component of the magnetization to
\begin{equation}
\omega/2\pi= J+\overline{D}/2.
\end{equation}
On the other hand, interestingly, the $I_zS_z$ term results in an enhancement of the signal associated with antisymmetric $J-$coupling as illustrated in Fig.\,\ref{Fig:Evolution}D.
Moreover, the PNC-induced y-magnetization is no longer static. 
Because the singlet spin state commutes with $I_z S_z$, the induced evolution must involve only the triplet states, so 
it oscillates at frequency $3\overline{D}/4$, corresponding to the average energy difference between the $|T_0\rangle$ and  $|T_{\pm1}\rangle$ states.
Importantly, this frequency is distinct from the frequency with which the electric field is modulated ($J+\overline{D}/2$), provided that $J\neq \overline{D}/4$.

\section{The proposed experiment}
\label{Sec:ProposedExpt}
Our proposed experiment involves excitation of magnetization along the y-axis. An experimentally realizable initial condition 
is:
\begin{multline}
{\rho}(0)=\frac{1}{4}
+\frac{B_\mathrm{p}}{4kT}\left[\frac{\gamma_I
+\gamma_S}{2}({{I}}_{x}+{{S}}_{x})
\right.
\\\left.+\frac{\gamma_I-\gamma_S}{2}({{I}}_{x}-{{S}}_{x})\right],
\end{multline}
which corresponds to magnetizing the spins along the x-axis in a field of strength $B_\mathrm{p}$ (where $|\gamma B_\mathrm{p}|\gg |2\pi J|$) at a temperature $T$; $k$ is Boltzmann's constant.
Following prepolarization, the sample is transported to the detection region in the presence of a guiding magnetic field, $S$ is inverted, and then, the guiding field is suddenly turned off. The inversion of $S$ can be done either in high field \cite{Tayler2016} or at zero field \cite{Min2018}; the purpose of the inversion is to maximize the ${{I}}_{x}-{{S}}_{x}$ and, correspondingly, the useful signal.

When $J_\mathrm{PNC}\neq0$, there is nonzero magnetization along $\hat{y}$ that oscillates at frequency $3 \overline{D}/4$ (see Fig.\,\ref{Fig:Evolution}D). 

Figure \ref{Fig:Evolution} shows a simulation of the PNC-dependent signal for $^{1}$H$^{19}$F. 
We assume the spins are prepolarized in a field $B_p=20$~T at 300\,K, and that the spin coherence time is 1\,s.
The simulated sample is $10^{21}$ molecules (50 $\mathrm{\mu L}$) at a distance of 7 mm from the magnetometer cell, which is typical of zero-field NMR detection. 
Figure \ref{Fig:Evolution}D shows the time evolution of the magnetic field at the magnetometer cell, which has an amplitude of $\approx 1.5\times10^{-16}$~T. Given a magnetometer sensitivity of $10^{-14} ~\mathrm{T}/\sqrt{\mathrm{Hz}}$ and a realistic duty cycle of 10\%, this would require on the order of hours to achieve a signal-to-noise ratio greater than unity. 
We emphasize that this level of signal corresponds to readily accessible laboratory conditions similar to those in zero-field spectrometers currently in use \cite{Blanchard2017,Tayler2017,CASPEr-Comagnetometer}. 

Note the beating in the in $\hat{x}$-signal (Fig.\,\ref{Fig:Evolution}D) that does not appear in the parity-violation related $\hat{y}$-signal. 
The origin of the beating is that the $\hat{x}$-signal has a component at a second frequency corresponding to the $|S_0\rangle \leftrightarrow |T_{\pm1}\rangle$ energy interval, $J-\overline{D}/4$.
This component arises from the $I_x-S_x$ component of the initial state. Because the $\hat{x}$-component is several orders of magnitude larger than the $\hat{y}$-component, in a realistic experiment, there will be inevitable ``leakage'' of the   $\hat{x}$-signal into the $\hat{y}$-channel. However, the distinct time dependence of the former will allow one to subtract (i.e., ``fit out'') the accurately measured $\hat{x}$-signal from the $\hat{y}$-channel, facilitating the discrimination of the sought-after signal.

\subsection{Systematic effects}

\begin{figure}
    \centering
    \includegraphics[width=\columnwidth]{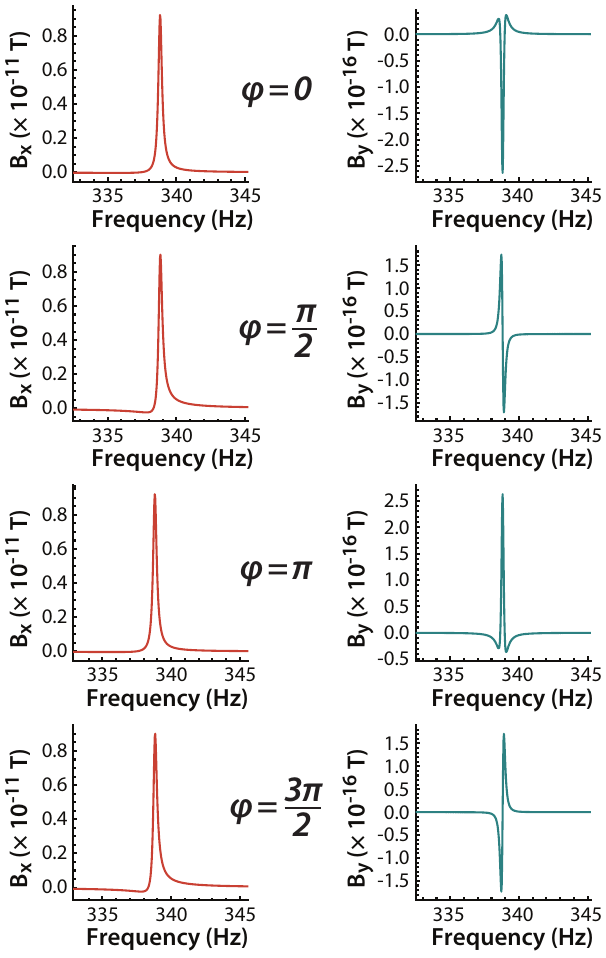}
    \caption{\textbf{Electric field phase cycling:}
    Dependence of $B_x$ (parity-conserving) and $B_y$ (PNC) signals on the phase of the electric field modulation.
    }
    \label{Fig:PhaseCycling}
\end{figure}
 
Systematic errors can be addressed by considering how they are affected by reversals of the electric field $\bm{E}$, the initial nuclear magnetization $\bm{M}_0$, and the sensitive axis of the detector, $\bm{B}_{Det}$, considering that the PNC signal transforms as the rotational invariant of Eq.\,\eqref{Eq:RotInvariant}.
 

Our method has the advantage that the PNC signal will emerge at the frequency $3 \overline{D}/4$, which is different from that of the driving electric field ($J + \overline{D}/2$). 
However, a component of the parity conserving NMR signal also evolves at this frequency, with amplitude orders of magnitude larger than that of the PNC signal, and it is likely impossible to avoid slight misalignments that would result in the magnetometer signal having at least some component of the parity conserving signal added to the PNC signal.
Fortunately, the parity conserving and nonconserving signals depend differently on the phase of the oscillating electric field.
Indeed, the phase of the PNC signal depends linearly on the phase of the electric field, while 
the parity conserving signal is invariant under reversal of the phase (see Fig.\,\ref{Fig:PhaseCycling}).
The PNC signal can thus be isolated by performing subsequent experiments with the electric field orientation reversed (a $180^\circ$ phase shift), and subtracting the signals.

In the design of the experiments, in addition to misalignment, one needs to consider also the effect of the ac magnetic field $\bm{B}_{ac}$ that would inevitably accompany the oscillating electric field. 
One example of potentially dangerous systematic corresponds to the rotational invariant $\bm{B}_{Det}\cdot \bm{B}_{ac}$. 
Indeed, this signal reverses with the reversal of the applied electric field (which is the origin of $\bm{B}_{ac}$) and the detection direction. 
However, this particular effect can be discriminated based on the fact that it does not reverse with the initial magnetization of the sample.

Other potential systematic effects may arise due to the presence of stray magnetic fields.
Perhaps the most concerning issue is that an ac magnetic field at the frequency of the electric-field modulation and directed along the $\hat{z}$-axis has a very similar effect to that of $\overline{J_{\rm{PNC}}}$, generating $I_y+S_y$ spin order, and thus an oscillating magnetization along $\hat{y}$ at the frequency $3\overline{D}/4$.
For a $^1$H--${}^{19}$F spin system, $B_z(t)\approx 100$\,fT yields an effect of the same magnitude as $\overline{\bm{J}^{(1)}}$ under the conditions discussed above.
This stray-field-induced systematic signal transforms as the rotational invariant 
\begin{equation}
    \bm{B}_{Det}\cdot \bm{M}_0\times \bm{\dot{B}}_{Str},\label{Eq:RotInvariantBSys}
\end{equation}
which is similar to Eq.\,\eqref{Eq:RotInvariant}, but with $\bm{E}$ replaced by $\bm{\dot{B}}_{Str}$.
This systematic effect can be monitored by sensitive magnetometry and discriminated, for instance, by its relative phase.


\section{Conclusions}

In conclusion, the nuclear spin dependent parity nonconserving contribution to the \textit{J}-coupling Hamiltonian in diatomic molecules puts observation of molecular PNC within the reach of current atomic vapor-cell magnetometers, even for molecules containing relatively light atoms such as HF. A detailed analysis of systematic effects will need to be carried out in conjunction with a specific experimental design; many potentially dangerous systematic effects can be discriminated based on several available reversals: that of the initial magnetization, detection direction, and and the phase of the applied electric field modulation. Additional tell-tale features of the useful signal include the fact that the signal appears at at frequency different from the frequency with which the electric field is modulated; moreover the PNC-signal frequency scales as the square of the amplitude of the applied field [see Eq.\,\eqref{Eq:AverageD}].

For simplicity we considered diatomic molecules as an example. However, this method should work also for more complicated non-chiral polar molecules. The size of the respective PNC effect can be estimated using Eq.\,\eqref{Eq:final_PNC}. For molecules with spin-1/2 nuclei in the range $7\le Z\le 81$, the PNC coupling is likely to be between a few $\mu$Hz and a few mHz as suggested by the estimates \eqref{estimate_HF} and \eqref{estimate_TlF}.


\section*{Acknowledgements}
This work was supported by the National Science Foundation under grant CHE-1709944 and by the Cluster of Excellence ``Precision Physics, Fundamental Interactions, and Structure of Matter'' (PRISMA+ EXC 2118/1) funded by the German Research Foundation (DFG) within the German Excellence Strategy (Project ID 39083149). M.K. is grateful to the Mainz Institute for Theoretical Physics (MITP) for its hospitality and acknowledges support from Russian Science Foundation under Grant No.\ 19-12-00157. The authors wish to acknowledge Prof.\,Robert Harris for useful discussions on symmetry and parity nonconservation, Prof.\,David DeMille for his comments on the theory and possible experimental implementations, and Dr.\ Leonid Skripnikov for finding a numerical error in our estimates.


\bibliography{PNC}


\beginsupplement
\section*{Supplemental Material}
\subsection{Symmetry Rules for Rank-1 \textit{J}-Couplings 
}\label{Supp-symmetry}

Since we are considering molecules oriented in an external electric field, the relevant parameter in our proposed experiments is the projection of $\bm{J}^{(1)}$ along the molecular electric dipole vector $\bm D$, or nonzero scalar product $\bm{J}^{(1)}\cdot \bm D$. 
According to Eq.\,\eqref{rank1}, the parity-conserving part of the rank-one $J$-coupling corresponds to an axial pseudovector $\bm{J}^{(1)}_\mathrm{PC}$.  
Consequently, the product $\bm{J}^{(1)}_\mathrm{PC}\cdot \bm D$ is a pseudoscalar. 
This product is therefore forbidden for molecules which have improper symmetry elements (improper rotations, reflections, and inversion), because a pseudoscalar changes sign under any such operation. 
We conclude that parity-conserving vector coupling can have nonzero projection on the molecular dipole moment only for chiral molecules. On the other hand, as one can see from Eq.\,\eqref{mixed-PNCa}],  parity violating  $\bm{J}^{(1)}_\mathrm{PNC}\cdot \bm D$ is nonzero even for diatomic polar molecules.

\subsection{Relativistic Corrections}

Equation \eqref{J_estimate} assumes the non-relativistic limit, but for heavy atoms relativistic effects may be important. The relativistic analogue of Eq.\ \eqref{NSD-PNC} is \cite{Bouchiat1974,Khr91}
 \begin{align}\label{NSD-PNC-rel}
 {H}_\mathrm{PNC}
 &=\frac{G}{\sqrt2}
 \sum_{i,K} g_K \bm \alpha_i{\bm I}_K
 \rho_K(\bm r_i-\bm R_K)\,,
 \end{align}
where $\bm \alpha$ is the Dirac matrix and $\rho_K(\bm r)$ is the
density of the valence nucleon in the nucleus $K$. This operator
does not contain the electron momentum $\bm p$ and is not changed in an
external magnetic field. Therefore, there is no
first-order contribution to the \textit{J}-coupling and we need to calculate the
second-order cross term between $ H_\mathrm{PNC}$ and the parity-even magnetic hyperfine interaction \cite{Sob79}:
 \begin{align}\label{HF-rel}
 {H}_\mathrm{HF}
 &= \sum_{i,K}
 \gamma_K \frac{{\bm I}_K \cdot\bm \alpha_i \times (\bm r_i -\bm R_K)}
 {(\bm r_i -\bm R_K)^3}\,.
 \end{align}

The second-order expression for the $J^{(1)}$ has the form:
 \begin{align}\label{J_second}
 J^{(1)}
 \sim
 \sum_n
 \frac{\langle 0 |H_\mathrm{PNC}|n \rangle
       \langle n|H_\mathrm{HF}|0 \rangle}
      {E_0-E_n}
 \,,
 \end{align}
where the sum over intermediate states includes both positive (electron) and negative (positron) energy states. In the sum over negative energy states we can approximately substitute the energy denominators with a constant $2m_e c^2$ and after that use the closure relation \cite{ASVJ99}. This leads us to the effective operator analogous to the second term in Eq.\,\eqref{NSD-PNC-A} which is bilinear in nuclear spin operators and where the $\delta$-function is replaced by the nuclear density $\rho_K$. For heavy atoms, the electron density at the origin $\rho_e$ is enhanced over the non-relativistic approximation \eqref{e-density} by a factor of \cite{Khr91}:
 \begin{align}\label{RelEnhancement2}
 &F_\mathrm{rel}
 = \frac{2(1+\gamma)(2ZR_\mathrm{nuc})^{2\gamma-2}}
 {\Gamma^2(2\gamma+1)}\,,
 \,\,
 \gamma=\sqrt{1-(\alpha Z)^2}.
 \end{align}
Here $\Gamma(x)$ is the gamma-function and $R_\mathrm{nuc}$ is the nuclear radius in atomic units. This factor is close to unity for $Z\lesssim 20$ and then grows rapidly with $Z$. For example, for the  $^{88}$Sr$^{38}$, $^{137}$Ba$^{56}$, $^{203}$Tl$^{81}$, and $^{226}$Ra$^{88}$ nuclei,  $F_\mathrm{rel}=1.64,\, 2.75,\, 7.64$, and $10.9$, respectively. Adding the factor \eqref{RelEnhancement2} to the expression \eqref{mixed-PNCb} we get Eq.\,\eqref{Eq:final_PNC}.

Up to now we have considered only the negative energy states in the sum \eqref{J_second}. In the sum over positive energy states the energy denominator is large for the core shells and the dominant contribution comes from the valence electrons. For this case the matrix elements of the PNC and HF interactions can be estimated as (see, for example, \cite{Khr91,Sob79}):
 \begin{align}
 \label{PNC_val}
 H_\mathrm{PNC}
 &\sim
 \frac{G\alpha Z^2}{2\sqrt2\pi}
 g
 \,,
 &H_\mathrm{HF}
 &\sim
 \alpha^2\frac{m_e}{m_p}Z
 \,.
 \end{align}
Assuming that typical energy denominators in \eqref{J_second} are of
the order of a Rydberg, we get an estimate:
 \begin{align}\label{J_estimate2}
 J^{(1)}
 \sim
 \frac{G\alpha^3 Z_I^2 Z_K}{2\sqrt2\pi}
 \frac{m_e}{m_p} g_I
 \,,
 \end{align}
where $Z_I$ and $Z_K$ are the two heaviest nuclei in the molecule and we assumed that $Z_I > Z_K$. Comparing this estimate with \eqref{J_estimate} we see that the main difference is substitution $Z_I^3 \to Z_I^2 Z_K$. If $Z_I\approx Z_K$ these estimates are close, but for the case $Z_I\gg Z_K$ the first-order contribution \eqref{J_estimate} dominates. We conclude that if the molecule has only one heavy nucleus, we can use expression \eqref{Eq:final_PNC}. Otherwise it is necessary to add the positive energy contribution, which requires elaborate numerical calculations.

\end{document}